# Fast Beam Condition Monitor for CMS: performance and upgrade

Jessica L. Leonard [a] *, Alan Bell [a], Piotr Burtowy [b], Anne Dabrowski [c], Maria Hempel [a,d], Hans Henschel [a], Wolfgang Lange [a], Wolfgang Lohmann [a,d], Nathaniel Odell [e], Marek Penno [a], Brian Pollack [e], Dominik Przyborowski [f], Vladimir Ryjov [c], David Stickland [g], Roberval Walsh [h], Weronika Warzycha [i], Agnieszka Zagozdzinska [j]

[a] DESY, 15738 Zeuthen, Germany
[b] Gdansk University of Technology, 80-233 Gdansk, Poland
[c] CERN, 1211 Geneva 23, Switzerland
[d] Brandenburg Technical University, 03046 Cottbus, Germany
[e] Northwestern University, Evanston, IL, 60208, USA
[f] AGH University of Science and Technology, 30-059 Krakow, Poland
[g] Princeton University, Princeton, NJ, 08540, USA
[h] DESY, 22607 Hamburg, Germany
[i] University of Warsaw, 00-927 Warsaw, Poland
[j] Warsaw University of Technology, 00-661 Warsaw, Poland

* Corresponding author. Contact information: Address: Jessica Lynn Leonard, DESY, Platanenallee 6, 15738 Zeuthen, Germany.  E-mail address: jessica.lynn.leonard@desy.de  Phone number: +49 40 8998 7391.

**Abstract**

The CMS beam and radiation monitoring subsystem BCM1F (Fast Beam Condition Monitor) consists of 8 individual diamond sensors situated around the beam pipe within the pixel detector volume, for the purpose of fast bunch-by-bunch monitoring of beam background and collision products. In addition, effort is ongoing to use BCM1F as an online luminosity monitor. BCM1F will be running whenever there is beam in LHC, and its data acquisition is independent from the data acquisition of the CMS detector, hence it delivers luminosity even when CMS is not taking data. A report is given on the performance of BCM1F during LHC run I, including results of the van der Meer scan and on-line luminosity monitoring done in 2012. In order to match the requirements due to higher luminosity and 25 ns bunch spacing, several changes to the system must be implemented during the upcoming shutdown, including upgraded electronics and precise gain monitoring. First results from Run II preparation are shown.

## 1. Introduction

*1.1 The Large Hadron Collider and Compact Muon Solenoid*

The Large Hadron Collider (LHC) [1] is a 27-km-circumference circular accelerator located at CERN near Geneva, Switzerland.  Up until 2013 it has collided beams of protons and lead ions at center-of-mass energies up to 8 TeV (2.76 TeV per nucleon pair for lead ions) with a peak instantaneous luminosity near $5 \times 10^{33}$ cm$^{-2}$ s$^{-1}$.  The Compact Muon Solenoid (CMS) [2] detector is one of four large detectors situated around the LHC and one of two general-purpose detectors.  CMS uses a layered barrel structure to distinguish the collision products and record their energy and momenta.

*1.2 Beam Condition Monitoring*

The CMS Beam Condition Monitoring (BCM) system serves to monitor the condition of the LHC beams at CMS in real time. Beam background is produced by proton interactions with the collimators or residual gas in the vacuum chamber. The beam background particles travel with the beam and enter the CMS detector, generating hits in the subdetector components. The BCM system gives an indication of the inner detector occupancy and therefore is used to ensure sufficiently low occupancy for data-taking. In addition, due to adverse conditions the LHC beam may lose some of its highly energetic particles into the area surrounding the beam pipe. Because the LHC runs at unprecedented intensities and beam energies, even small beam losses may cause damage to CMS detector components. The BCM system therefore monitors the particle flux around the beam pipe for beam loss conditions and can initiate fast reactions when necessary, such as a beam abort.

*1.3 Fast Beam Condition Monitor BCM1F*

The CMS Fast Beam Condition Monitor (BCM1F) [3] records bunch-by-bunch information via particle hits with a resolution better than the time between bunch crossings. Since it is able to detect beam halo as well as collision products, it is being used to provide values for both beam background and luminosity. Its readout is independent of CMS data acquisition, since the background and luminosity should be monitored even when CMS is not running.

BCM1F consists of 8 single-crystal chemical vapor deposition (sCVD) diamond sensors, each 5 mm x 5 mm and 500 μm thick. The sensors are located at a radial distance of 5.5 cm from the beam pipe and are arranged in two parallel rings at 1.8 m to either side of the interaction point. Diamond is used because of its radiation-hardness, its high signal generation, and its lack of need for cooling. The diamond sensors are metallized and integrated into frontend modules containing a radiation-hard preamplifier/shaper and an optical driver. The optical driver transmits the pulses via fiber to an optical receiver in the counting room, where the information is fanned out into parallel paths. The primary data-taking path consists of a fixed-threshold discriminator whose output rates have been read out alternately with a time-to-digital converter, a scaler, and a custom histogramming board (see section 3.3). Additional logic modules include a lookup table, which generates coincidences of sensor hits, and a multiple gate and delay module, which generates vetoes and gates on-the-fly for selecting specific points in the orbit structure. The secondary path, consisting of an analog-to-digital converter (ADC), records the full transmitted signal over an orbit and is used primarily for monitoring and efficiency calculations. The ADC has a large deadtime which has prevented this path from being used for data acquisition.

The position of the BCM1F sensor planes along the beam pipe is optimal for separation of background and collision products, which arrive at a given plane 12.5 ns apart. Incoming beam brings with it beam halo products only, while the outgoing beam is accompanied by beam halo products as well as collision products. Therefore to measure beam background, only incoming beam must be considered, while the outgoing beam is useful for the luminosity calculation because the collision products outnumber the background by a factor of $10^5$. Still, because BCM1F has such a small sensor surface area, the fraction of bunches that are actually detected is small.

**2. BCM1F Performance as a Luminometer**

The BCM1F hit rates have been measured to be linear with luminosity as shown in Figure 1, making them a good observable with which to calculate luminosity. The threshold for hit detection is set higher than the noise level, so noise does not make a contribution to the hit rate. However, albedo (secondary interactions) have been measured to contribute at a level of 1% of the total hit rate. The non-trivial part is the calibration of the measurement. This is done with a van der Meer scan, in which the two beams are scanned across each other in steps in both the *x* and *y* directions, and the hit probability is measured at each step. Since the number of observed interactions is assumed to be a Poisson distribution, where the probability of detecting n interactions is given by $P_n = \mu^n e^{-\mu}/n!$, the most probable value $\mu = -ln[P_0]$ can be found by measuring the probability of detecting zero hits $P_0$. $\mu$ is plotted against beam separation distance for both the *x* and *y* directions, and the resulting distributions are approximately Gaussian. The luminosity can be calculated from the equation $L = \mu_{vis} n_b f_{orbit}/\sigma_{vis}$, where $\mu_{vis}$ is the average number of interactions detected, $n_b$ is the number of colliding bunches, $f_{orbit}$ is the orbit frequency, and $\sigma_{vis}$ is the visible interaction cross section. This expression can also be written as $L = f_{orbit} N_1 N_2 / 2\pi \Sigma_x \Sigma_y$, where $N_1$ and $N_2$ are the numbers of particles per bunch in beam 1 and beam 2 respectively, and $\Sigma_x$ and $\Sigma_y$ are the widths of the $\mu$ distribution plotted with respect to *x* and *y* beam separation as determined from the van der Meer scan. An example fit to find the quantity $\Sigma$ is shown in Figure 2. This calibration constant was calculated for each of the bunches in the orbit and compared to that calculated by the other luminosity subdetectors. As shown in Figure 3, BCM1F agrees favorably with the other subdetectors. The average value calculated over the full orbit agrees to the other subdetectors to within 1%.

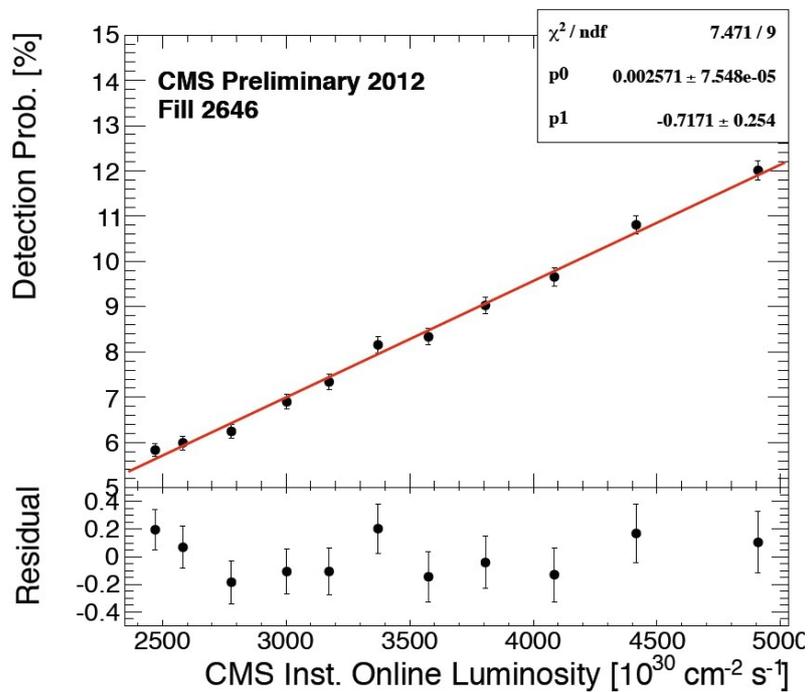

*Figure 1. BCM1F hit rate as a function of instantaneous luminosity. The relationship between hit rate and luminosity is highly linear.*

## 3. BCM1F Upgrade

*3.1 Overview*

The LHC upgrade has several implications for the BCM1F system that need to be addressed. During Run I the characteristic instantaneous luminosity (5 x $10^{33}$ cm$^{-2}$s$^{-1}$) corresponded to a hit rate of 3.7 x $10^6$ s$^{-1}$ per sensor (1.5 x $10^7$ cm$^{-2}$s$^{-1}$) and a sensor occupancy of ~12%. For Run II the instantaneous luminosity is foreseen to increase from the previous maximum of ~7 x $10^{33}$ cm$^{-2}$s$^{-1}$ to around $10^{34}$ cm$^{-2}$s$^{-1}$, corresponding to 7.5 x $10^6$ s$^{-1}$ per sensor (3 x $10^7$ cm$^{-2}$s$^{-1}$) and a sensor occupancy of ~25%. The higher radiation field means that the BCM1F design and the components that sit close to the beam pipe should be optimized for radiation exposure as well as the higher hit occupancy per bunch crossing. The new system will need to be able to measure the high rates for collisions in addition to the low rates from the beam background, which occur at a level $10^5$ lower than collision rates. Furthermore, the LHC will potentially run with a bunch interval of 25 ns, half that used during Run I. The old system's electronics need to be updated to handle the smaller bunch spacing. The post-upgrade LHC running conditions require many changes to be made to the various components of the BCM1F system. Specific details will be given in the following sections.

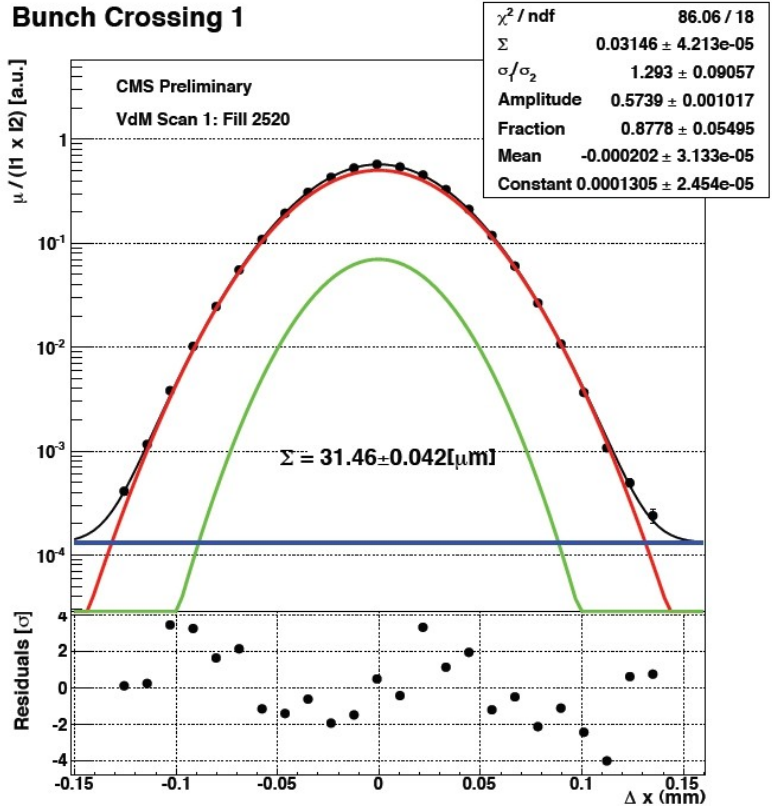

*Figure 3. Example fit to determine luminosity calibration. The average number of proton interactions μ is plotted against the beam separation distance. The width of the resulting distribution provides the calibration constant.*

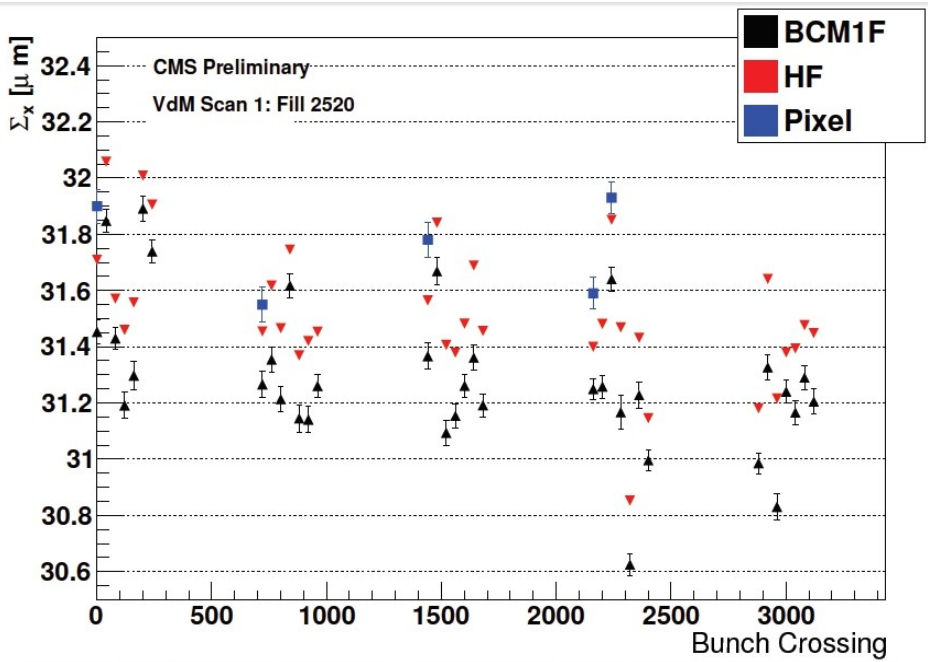

*Figure 2. BCM1F luminosity calibration constant compared with other subdetectors. The bunch-by-bunch calibration constant for the x direction $\Sigma_x$ as measured with BCM1F is compared with that measured by other luminosity subdetectors over a full orbit of a van der Meer scan, with agreement of the average values to within 1%.*

*3.2 Carriage, Sensors, and Frontend Electronics*

The new concept for BCM1F has a much higher number of channels. The 8 diamonds of the old system will be replaced by 24: 12 diamonds in two parallel planes on either side of the interaction point. Additionally, each diamond will be split into two channels by means of a split metallization. Having more channels makes the luminosity relationship to pileup more linear. To extend to higher luminosity (and therefore pileup) values, having more channels makes it more likely that at least one sensor will have zero hits per bunch crossing, i.e. gives a higher probability of not reaching a saturation state. For the luminosity calculation, the number of "zero" or empty occurrences in the observable used for calculation (sensors, bunch buckets) is the important quantity. Having more channels also makes it more likely that at least one sensor will register a beam halo hit at the incoming beam. An example placement of the diamonds in one plane with the double metallization is shown in Figure 4. A new carbon-fiber carriage will hold the increased number of BCM1F sensors as well as the frontend electronics.

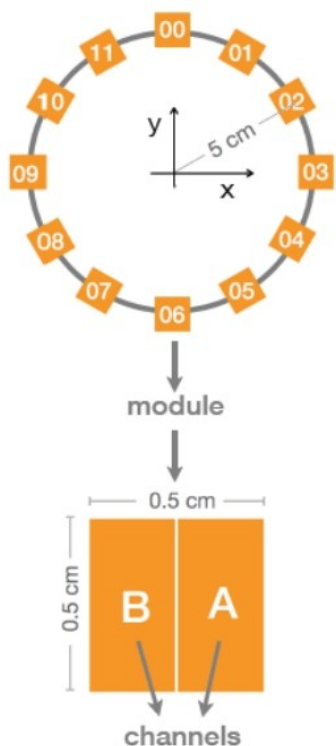

*Figure 4. Diagram of possible BCM1F upgrade diamond placement. 12 diamond sensors are arranged around the beam pipe with a uniform radius. The split metallization forming two channels per sensor is also shown.*

The new system will continue to use sCVD diamond sensors because of their radiation-hardness, good charge signal, and lack of cooling needs. However, despite diamond's inherent resistance to radiation damage, some damage was seen during Run I in the form of polarization. Ionizing radiation causes charge traps in the material, and when subject to both radiation and a voltage difference (as during running), the charge traps fill preferentially according to the electric field. The trapped charge compensates for the electric field in the bulk of the material, which serves to reduce the charge collection efficiency of the sensor. The radiation exposure of the sensors cannot be prevented, but strategies can be adopted to deal with the situation. The current primary line of defense against polarization is to use a higher voltage across the sensor; this improved performance of the damaged sensors during Run I. Additional ideas under consideration include illumination with red light or alternating the polarity of the high voltage, both of which would free the trapped charge, although these would be more difficult to implement and are therefore disfavored.

The frontend ASIC used during Run I had several characteristics that decreased the efficiency of the system. A long rise time of 25 ns prevented immediately-subsequent signals from being seen. Other effects were only relevant in the rare cases when unusually large amounts of charge were deposited in the sensors. A large signal would saturate the amplifier for a time on the order of 100 ns, causing many bunch crossings' worth of channel deadtime. When the amplifier came out of its saturation state, it would "overshoot" and return to a state much beyond the nominal baseline for a time period of up to a few μs. During this time any signal pulses would be transmitted but may not have exceeded the discriminator threshold in order to be counted. In this way the large pulses caused significant deadtime.

To deal with these situations and improve the timing resolution of BCM1F for the shorter bunch intervals, a new fast radiation-resistant ASIC using 130 nm technology was developed by AGH in Krakow. It was designed to have a much shorter rise time and a much smaller reaction to large signals. Preliminary tests have confirmed this behavior. Figure 5 shows the typical signal, with a rise time of about 7 ns and a pulse width of about 8 ns. Figure 6 shows the response to large charge deposits: for a signal of 150 fC, the total time-over-threshold is about 30 ns, with an extremely small overshoot time if any. These results show a significant improvement over the old system and as such indicate a good prognosis for the post-upgrade system.

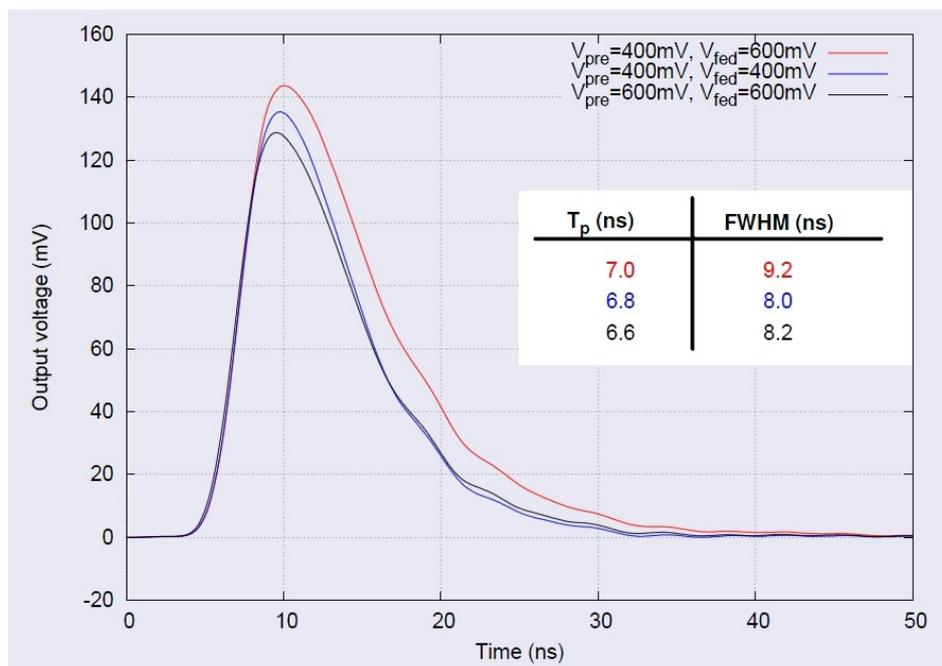

*Figure 5. Typical signal shape for BCM1F upgrade frontend ASIC. The signal has a rise time of about 7 ns and a full-width half-maximum value of 8-9 ns.*

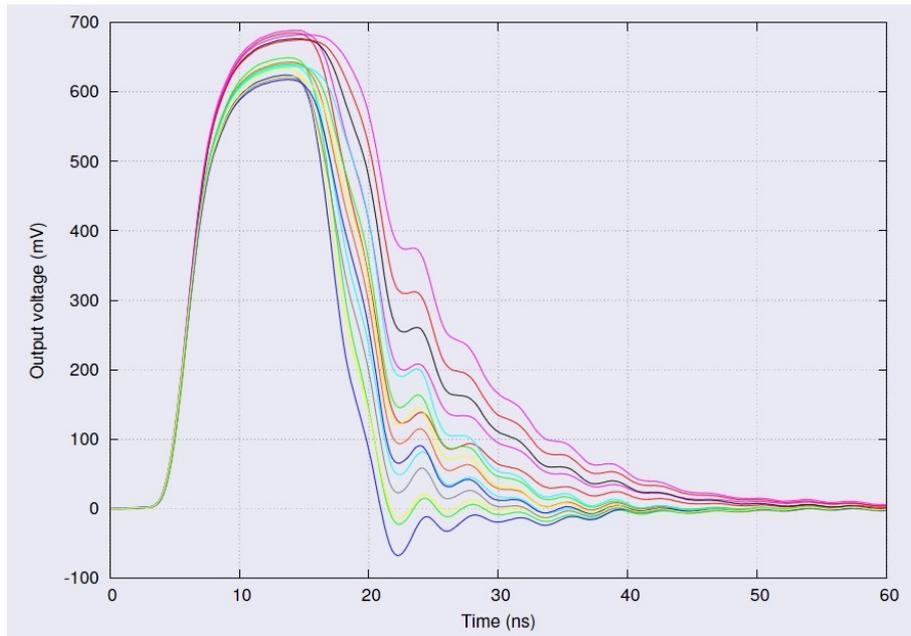

*Figure 6. Behavior of BCM1F upgrade frontend ASIC for large charge deposits. Maximum time-over-threshold is around 30 ns, and the time spent in overshoot is extremely short.*

The optical driver transmitting the signal to the backend electronics also suffered the effects of radiation damage. Over the course of Run I (30 fb$^{-1}$) 25% of the optical gain was lost as measured by the baseline position and response to test pulses injected into the electronics. For the upgrade the optical driver will be moved from the curved part of the carriage to the long arm of the carriage, out of the strongest radiation field, in order to reduce its radiation exposure. The new location of the driver is shown in Figure 7. In addition, the test pulse system will use multiple amplitudes to monitor the threshold and linearity of the laser response over time. Furthermore, since the optical driver was found to be sensitive to temperature, a fiber Bragg grating temperature sensor will be used to monitor and enable corrections for the temperature. BCM1F will also benefit from temperature stabilization systems being installed by nearby subdetectors.

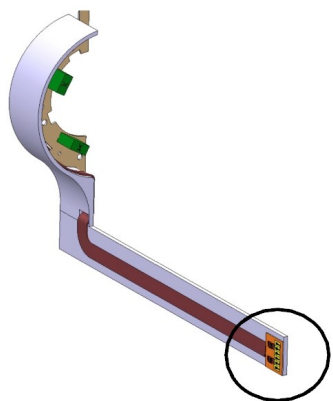

*Figure 7. Laser driver position on carriage for BCM1F upgrade. The new position of the laser driver ensures a lower radiation dose to maintain performance of the optical chain.*

*3.3 Backend Electronics, Data Acquisition, and Luminosity Calculation*

The parallel-path structure of the BCM1F backend will be updated for the upgrade running. For the discriminator path, several new discriminators have been considered, with the main choices being the CAEN V895 (a fixed-threshold discriminator) and the PSI CFD950 (a constant-fraction discriminator). Although the constant-fraction discriminator has better single-pulse time resolution, the fixed-threshold discriminator has a lower deadtime, and it was determined during tests that the deadtime had a larger effect than the single-pulse resolution. Therefore the fixed-threshold discriminator is the first choice for installation, although it is planned to run a constant-fraction discriminator in parallel for development purposes.

In addition, the ADC path will be updated. In the old system the ADC was used only for monitoring, although with advances in hardware, it is foreseen to possibly use a fast ADC digitizer in parallel for future full-time data-taking. For this purpose multiple peak-finding and deconvolution algorithms are being investigated. In addition to determining a given pulse's arrival time, the digitizer would also be able to measure the signal height, detecting multiple simultaneous MIPs, as well as distinguish between signals that are overlapping in time, providing an advantage over threshold-based discrimination. Detected hits will be histogrammed with 4 bins per bunch crossing. Investigation into hardware options is ongoing. Currently the FMC125 ADC from 4DSP is being tested for this purpose. It can sample 4 channels with 1.25 GS/s per channel at 8-bit resolution.

It is foreseen to use the already-tested discriminator path for initial running, while commissioning the digitizer path in the meantime. If the digitizer path is fully validated with no deadtime, then it may be turned to as the primary peak-finding and data acquisition method.

A dedicated data acquisition board has been developed at DESY-Zeuthen for BCM1F readout, the Real-time Histogramming Unit (RHU). The RHU accumulates 8 ECL input channels of full-orbit histograms with bins of 6.25 ns (4 bins per bunch bucket) for a total of 14256 bins per orbit. The data acquisition and readout is deadtimeless due to the use of a double buffer system. The unit also takes several NIM control signals as input: LHC bunch clock, orbit clock, and beam abort. The interval of histogram accumulation is configurable, and the board is read out via Ethernet.

The data acquisition makes use of a shared memory so that multiple applications are able to access the data simultaneously. The data is transmitted by a DAQ sender process running on the RHU's on-board Linux system to a provider process on the DAQ computer, which puts the data into a shared memory. Transmission to multiple DAQ computers is supported. Individual client applications (such as real-time display or file storage) running on the DAQ computer can retrieve the data from the shared memory, making use of the RHU software library. The software library and shared memory structure were designed to decouple the end-user software processes from the hardware as much as possible, increasing the flexibility and decreasing the complexity of the user software.

Several updates are planned for the next hardware revision of the RHU in order to improve flexibility and enable integration into a larger data acquisition system. The on-board input channels will be replaced by a mezzanine ECL connector, so that a standard 32-channel input mezzanine board can be used. This allows further (non-histogramming) input channels, as well as the possibility of custom mezzanine development. In addition, an optical fiber input will be added to receive external timing signals, namely an external signal to indicate the histogram sampling interval boundary.

The RHU will be integrated into the CMS luminosity data acquisition system LumiDAQ. A version of LumiDAQ existed during Run I, but the system is being expanded to include other luminosity detectors such as BCM1F. The purpose of LumiDAQ is to coordinate and combine the data from the individual CMS luminosity detectors. Data acquisition for each subsystem will be controlled by a

LumiDAQ master system, and data from each system will be transmitted to a single LumiDAQ resource broker. Since luminosity data will be combined from all luminosity subsystems, for transmission the individual systems must therefore use a common data format, chosen to be a histogram with one bin per bunch crossing (3564 bins per orbit). A conceptual diagram of how BCM1F fits into the LumiDAQ system is shown in Figure 8.

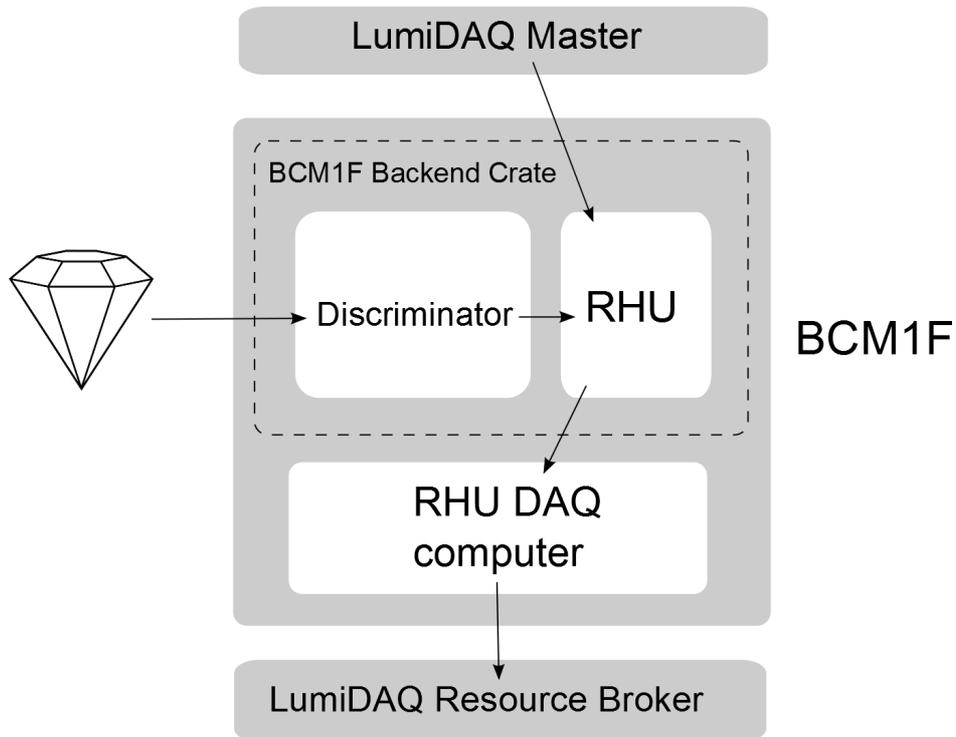

*Figure 8. Diagram of RHU integration with LumiDAQ system. The RHU will receive histogram input from the BCM1F discriminator and timing signals from the central timing system. It will send the output to be processed on the RHU DAQ computer. The data will then be passed to the LumiDAQ system, which will combine the luminosity information from all luminosity subsystems.*

The luminosity monitoring must take place even when CMS may not be taking data; therefore LumiDAQ runs will be decoupled from CMS runs. However, when CMS is running the data needs to be correlated with LumiDAQ, so the beginning of a CMS run always means the beginning of a new LumiDAQ run. The smaller intervals used for the calculation of instantaneous luminosity (known as "luminosity sections" and "luminosity nibbles") will be correlated in similar ways between CMS DAQ and LumiDAQ. Since multiple detectors will be providing data, they need to be very precisely aligned in terms of timing, necessitating the use of external timing signals. The timing signals will be distributed on optical fibers from the central CMS timing system. To maintain synchronization between the different systems in case one drops out for a period of time, counters will be distributed via the optical system which the subsystems will use to resynchronize themselves with LumiDAQ.

The algorithms for calculating luminosity also need to be updated. Pileup levels will be higher after the upgrade, potentially reaching over 100 before the end of Run II. The simple coincidence algorithm used during Run I is foreseen to saturate by pileup values of 30. Work is currently in progress to develop new algorithms that will provide responses useful at both high pileup and the low pileup values used during the luminosity calibration.

## 4. Conclusion

BCM1F has showed potential as an online luminometer in 2012 running. The luminosity calibration agreed on average with other subsystems' values to within 1%. In addition, the hit rate is linear over a range of luminosities, which indicates that a linear extrapolation to the post-upgrade period is reasonable.

There are many improvements in progress to increase the effectiveness of the BCM1F system post-upgrade. To deal with polarization due to radiation damage in the diamond sensors, it is planned to use higher voltage across the sensors to maintain charge collection efficiency. A new fast frontend ASIC has been developed to reduce inefficiencies related to a slow rise time and large charge deposits. Additionally, the laser drivers will be moved to a lower-radiation area, and a multi-amplitude test pulse will be used to monitor laser response. For the backend, a fixed-threshold discriminator will be used in parallel with a fast ADC digitizer. The Real-time Histogramming Unit was developed to provide full-orbit histograms with which to measure beam background and luminosity values. The RHU will be integrated into the already-existing LumiDAQ system, which combines the BCM1F values with other subsystems' values. In addition, algorithms for luminosity measurement in the post-upgrade conditions are currently under development. Current results of these developments give a good prognosis for BCM1F performance during Run II.